\documentclass[aps,prb,10pt,twocolumn,floatfix,showpacs,superscriptaddress]{revtex4-1}
\usepackage[colorlinks=true,citecolor=blue,linkcolor=blue,urlcolor=blue]{hyperref}
\usepackage{amsmath}
\usepackage{amssymb}
\usepackage{graphicx,amsmath,amsfonts,bm}
\usepackage{epstopdf}
\usepackage{bm}
\usepackage{bbm}
\usepackage{color}
\usepackage{relsize}
\usepackage{dsfont}
\usepackage{braket}
\usepackage[caption=false]{subfig}
\usepackage[all]{hypcap} 
\usepackage[T1]{fontenc}
\usepackage{dsfont}
\usepackage{mathbbol}
\usepackage{wasysym}
\usepackage{physics}
\usepackage{soul}
\usepackage{ulem}
\usepackage{hyperref}

\graphicspath{{Pictures/},{Pictures3/}}

\begin{document}
	\title{Interedge backscattering in quantum spin Hall-based NS and SNS junctions} 
	    \author{Cajetan Heinz}
	\affiliation{Institut f\"ur Mathematische Physik, Technische Universit\"at Braunschweig, D-38106 Braunschweig, Germany}
     \author{Patrik Recher}
	\affiliation{Institut f\"ur Mathematische Physik, Technische Universit\"at Braunschweig, D-38106 Braunschweig, Germany}
	\affiliation{Laboratory for Emerging Nanometrology Braunschweig, D-38106 Braunschweig, Germany}
	\author{Fernando \surname{Dominguez}}
	\affiliation{Institut f\"ur Mathematische Physik, Technische Universit\"at Braunschweig, D-38106 Braunschweig, Germany}
    \affiliation{Institute for Theoretical Physics and Astrophysics, and W\"urzburg-Dresden Cluster of Excellence on Complexity, Topology and Dynamics in Quantum Matter ctd.qmat, Julius-Maximilians-Universit\"at W\"urzburg, Am Hubland, D-97074 W\"urzburg, Germany}
	
	\date{\today}
 \begin{abstract}{
We investigate the microscopic conditions that allow for the coupling between opposite quantum spin Hall (QSH) edges in hybrid junctions with superconductors. Using a microscopic Bernevig--Hughes--Zhang model and the Bogoliubov--de Gennes formalism, we model a potential barrier along the NS interface and identify the parameter regimes in which the QSH edges are coupled. In normal--superconductor junctions, such coupling manifests as deviations from the quantized zero-bias Andreev conductance $G=4e^2/h$. These deviations are controlled by the induced gap in the barrier, the barrier geometry, the interface transparency, orbital and Fermi-velocity mismatch, and disorder strength as well as the bias voltage leading to a zero-bias peak. 
We then analyze the impact of this interedge-coupling
mechanism in Josephson junctions at equilibrium and show that it
hybridizes the edge-resolved Andreev branches, opens gaps at the time-reversal-invariant phase differences $\varphi=0$ and $\varphi=\pi$, and modifies the superconducting quantum interference pattern.
In a reflection-symmetric geometry, the relative sizes of the two gap openings provide complementary
information about the interedge dynamical phase, which also determines the parity of the suppressed lobes in the magnetic interference pattern.  This investigation sheds light on the microscopic details that control the coupling of helical edge states in actual devices and the resulting consequences for superconducting hybrid systems.}
\end{abstract}
\maketitle
\section{Introduction}
Topological superconductors can host Majorana bound states (MBS) at
boundaries and defects\cite{Volovik1999a, Read2000a, Kitaev2001a, Ivanov2001a}. These zero-energy modes are of considerable
interest because of their non-Abelian exchange properties and their
protection against local perturbations that do not close the bulk gap or
break the relevant protecting symmetries\cite{Ivanov2001a}. In this context, proximitized quantum spin Hall insulators (QSH)~\cite{Fu2009a} provide one of the simplest routes to topological superconductivity: the helical edge states supply an effectively spinless transport channel without the need for an external magnetic field.

In their seminal work, Fu and Kane~\cite{Fu2009a} showed that a Josephson
junction formed by two $s$-wave superconductors coupled through a single
QSH edge can host MBS at the QSH--superconductor interfaces. The
hybridization of these MBS across the weak link gives rise to a parity-resolved Andreev branch exhibiting 4$\pi$-periodic spectral flow. This is in
contrast to conventional Josephson junctions, where the ABS spectrum is
$2\pi$ periodic in the superconducting phase difference. In the ideal
parity-conserving limit, the resulting fractional Josephson effect can
manifest itself through missing odd Shapiro steps or through a halving of
the Josephson radiation frequency\cite{San-Jose2012a, Pikulin2012a, Dominguez2012a,Virtanen2013a, Houzet2013a, Dominguez2017a, Deacon2017a}. Despite substantial experimental
efforts~\cite{Hart2014, Pribiag2015a, Bocquillon2016a, Deacon2017a,
Bendias2018, Randle2023}, however, an unambiguous identification of MBS in
QSH-based Josephson junctions remains challenging.

\begin{figure}[t!]
  \centering    \includegraphics[width=\linewidth]{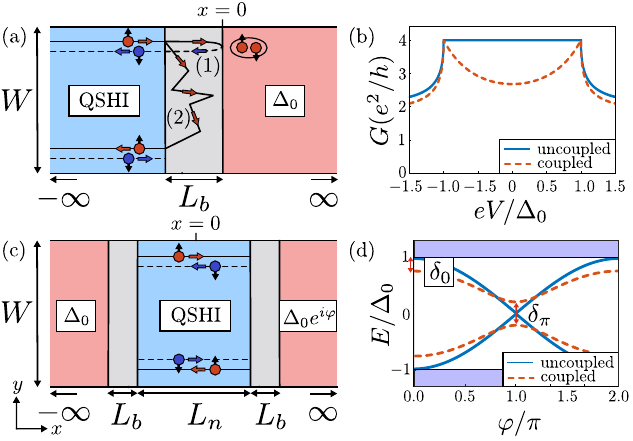}
  \caption{(a)~Schematic of a QSH/barrier/superconductor junction.
    In the barrier, electrons can undergo local Andreev reflection~(1)
    or undergo normal reflection to the opposite edge~(2). Red (blue) dots and solid (dashed) lines represent electrons (holes).
    (b)~Conductance $G$ as a function of bias voltage $eV$ for uncoupled
    (blue solid line) and coupled (red dashed line) top and bottom
    QSH edges. 
    (c)~Schematic of a superconductor/barrier/QSH/barrier/superconductor
    junction with phase difference $\varphi$ between the superconductors.
    (d)~Andreev bound state energies $E$ as a function of $\varphi$
    for uncoupled (blue solid line) and coupled (red dashed line) edges. 
  }
  \label{Fig1}
\end{figure}

One important obstacle is the limited microscopic control over the
superconductor--QSH interface. In particular, hybrid junctions consisting of a superconductor on top of a topological insulator can result in a poorly controlled interface, which may make transport signatures of topological superconductivity difficult to resolve~\cite{Pribiag2015a, Bocquillon2016a, Deacon2017a,
Bendias2018}. 
Alternative approaches involving a lateral attachment of the superconductor~\cite{Bai2020,Goldhaber-Gordon, Mandal2026AndreevCavity} can improve the superconductor--QSH coupling, however, even in this case, the band alignment and work-function mismatch between the QSH system and the superconductor can locally shift the chemical potential of the QSH edges near the interface and create a
conducting barrier, as illustrated in Fig.~\ref{Fig1}(a,c).
This conducting interface region can qualitatively modify the simple
Fu--Kane picture. In the idealized limit of well-separated edges, the top
and bottom QSH edges are effectively independent, and local
time-reversal-symmetric disorder cannot backscatter a particle within a
single helical edge. A conducting barrier changes this situation by
providing a nonlocal path through which a particle can propagate from one
edge to the other~\cite{Lee2014a,Baxevanis2015a}. Backscattering can then occur via the opposite edge without requiring magnetic perturbations or explicit time-reversal
symmetry breaking, yielding a reduction of the quantized conductance, see Fig.~\ref{Fig1}(b). 

Previous scattering descriptions~\cite{Baxevanis2015a} established that interedge coupling can modify the Andreev spectrum and produce an even--odd superconducting quantum interference (SQI) pattern, see Fig.~\ref{Fig1}(d). What remains less understood is how such coupling emerges from the microscopic properties of an extended QSH-superconductor interface and how its magnitude and phase depend on the geometry, band alignment, proximity effect, and disorder. These questions motivate the microscopic analysis developed below.

In this work, we use a microscopic tight-binding model to identify the
conditions under which the top and bottom QSH edges become coupled through a conducting interface region. We first study an NS junction and compute the zero-bias conductance as a diagnostic of interedge coupling. For two decoupled, perfectly Andreev-reflecting helical edges, the NS conductance takes the quantized value $G=4e^2/h$. Thus,  deviations from this value provide a practical way to map the parameter regimes in which the conducting barrier mediates scattering between opposite edges.

We then turn to SNS Josephson junctions in the intermediate junction regime $L\approx \xi_s$, where $L$ is the length of the normal region and $\xi_s$
is the superconducting coherence length. We show that the interedge coupling opens avoided crossings in the ABS spectrum at the time-reversal-invariant
phase differences $\varphi=0$ and $\varphi=\pi$, 
denoted by $\delta_0$ and $\delta_\pi$, which oscillate (out of phase) as functions of the
chemical potential and device geometry. In particular, we study their relation to the even-odd flux-quanta effect~\cite{Baxevanis2015a} observed in the superconducting quantum interference (SQI) pattern and its impact on the Fraunhofer pattern taking place when few additional bulk bands contribute to the supercurrent.

The paper is organized as follows. In Sec.~\ref{Microscopic Model}, we
introduce the BHZ model in Bogoliubov--de Gennes form and describe the
Green's-function transport formalism. In Sec.~\ref{NS junction}, we use
the NS conductance to identify the parameter regimes in which the two QSH
edges are effectively coupled. In Sec.~\ref{Andreev bound states and SQI pattern},
we analyze the consequences of interedge coupling for the ABS spectrum and
the SQI pattern in the intermediate-junction regime. We summarize our results in Sec.~\ref{sec:conclusions}.

\section{Microscopic Model}
\label{Microscopic Model}
In this section, we present the tight-binding formulation of the BHZ Hamiltonian along with the transport formalism used to compute the current and the conductance of the system.
\subsection{Tight-binding Hamiltonian}

We describe the NS and SNS junction using the BHZ model~\cite{Bernevig2006a} within the framework of the Bogoliubov-de Gennes formalism.  We neglect spin-nonconserving terms such as Rashba and bulk-inversion-asymmetry contributions, considered in Refs.~\onlinecite{Virtanen2012a,Haidekker2020a}. This approximation allows us to work in a reduced basis while preserving the helical edge structure relevant for the interedge-coupling mechanism. Spin-nonconserving terms can modify phases and spin textures, but they do not qualitatively change the nonlocal backscattering mechanism studied here
~\footnote{We consider deviations of our results produced by a finite SOC in App.~\ref{SOC and Fermi velocity mismatch}}.
Consequently, we can work in a reduced basis, reducing the computational costs of the numerical calculations. The Hamiltonian takes the form $H=(1/2)\int \text{d}x \text{d}y \Psi^{\dagger}\mathcal{H}\Psi$ with
\begin{align}
\label{BdG BHZ Hamiltonian}
    \mathcal{H}=\begin{pmatrix}
        \mathcal{H}_{e}(x)-E_{F} & \Delta(x) \mathbb{1}_{2\times 2}\\
        \Delta^{*}(x)\mathbb{1}_{2\times 2} & E_{F}-\mathcal{H}_{h}(x)
    \end{pmatrix},
\end{align}
where $E_{F}$ is the Fermi energy and $\mathcal{H}_{h}=\mathcal{T}\mathcal{H}_{e}\mathcal{T}^{-1}$ is the hole Hamiltonian, which is the time-reversed version of the electron Hamiltonian $\mathcal{H}_{e}$. In the full spinful BHZ basis, time reversal is represented by $\mathcal T=is_yK$. In the reduced BdG block used below, the spin reversal is implicit in the choice of the electron-up and hole-down sectors, so that the time-reversed orbital block is generated by complex conjugation. $\Delta(x)$ is the superconducting pairing which we model as
    $\Delta(x)=\Delta_0 \Theta(x)$
for the NS junction and as
\begin{align}
    \Delta(x)=&\Delta_0\left[\text{e}^{i\varphi}\theta\left(x-L/2\right)+\theta\left(-x-L/2\right)\right]
\end{align}
for the SNS junction, 
where $\Delta_0$ is the real amplitude of the pairing, $\varphi$ the superconducting phase difference between the left and right superconductor and $L=L_n+2L_b$ the length of the normal-conducting region, as indicated in Fig.~\ref{Fig1}(c).
We can express $\mathcal{H}$ in the reduced basis
\begin{align}
\label{Electron basis}
    \Psi(x,y)=\left[c_{E\uparrow}({\bf r}),c_{H\uparrow}({\bf r}),c_{E\downarrow}^{\dagger}({\bf r}),c_{H\downarrow}^{\dagger}({\bf r})\right]^{T}_{(x,y)},
\end{align}
for electrons (holes) with spin $\uparrow$($\downarrow$) in the orbital states $|E_1,\pm 1/2\rangle$ and $|H_1,\pm 3/2\rangle$, corresponding to the lowest electron-like and heavy hole-like subbands of the quantum well, respectively, at the spatial coordinates $(x,y)$.
The electronic Hamiltonian is given by
\begin{align}
\label{Electronic part} \mathcal{H}_{e}=&\varepsilon(\hat{k})+M(\hat{k})\sigma_{z}+A\left(\hat{k}_{x}\sigma_{x}-\hat{k}_{y}\sigma_{y}\right),
\end{align}
with $\varepsilon(\hat{k})=C(x)-D\hat{\textbf{k}}^2$, $M(\hat{k})=M(x)-B\hat{\textbf{k}}^2$ and $\hat{\textbf{k}}=-i \nabla_{\textbf{r}}$. The Pauli matrices $\sigma_{i}$ act in the pseudospin ($E_{1}$, $H_{1}$)-space. 
We model the mass in the different regions as $M(x)=M_{\text{QSH}}\theta(-x)+M_s\theta(x)$ in the NS junction and as 
\begin{align}
\nonumber
    M(x)=&M_s \theta(|x|-L/2)+M_{\text{QSH}}\theta(L/2-|x|)
\end{align} in the SNS junction,
with $M_s=10\,$meV as the mass in the superconducting region and  $M_{\text{QSH}}=-10\,$meV as the mass in the barrier and the QSH edges. Note that the change in sign in $M(x)$ when going from the QSH insulating to the superconducting regions accounts for the  ordinary character of the latter. In this way, the superconducting Hamiltonian covers the band structure of ordinary superconductors and proximitized HgTe in the highly doped limit~\cite{Lee2014a}. Similarly as for $M(x)$, we define a spatially varying potential $C(x)$ for the NS and SNS junctions. To this aim, we set $C_s$, for the superconducting sectors, $C_b$ for the barriers and $C_n$ for the rest of the QSH sector.

We use the following set of parameters: $A=373$ meV\,nm, $B=-857\,$meV\,nm$^2$, $D=-682\,$ meV\,nm$^2$ and $\Delta_0=0.15\,$meV. We set $E_{F}=0\,$meV and tune the chemical potential by the $C$ parameter. If not specified otherwise, we use $C_n=-4\,$meV, $C_s=20\,$meV and $C_b=19\,$meV, where the subscript indicates the normal region, superconductor or barrier, respectively.\\ We are interested in modeling a discrete two-dimensional material, therefore, we apply standard tight-binding discretization methods in Eq.~(\ref{BdG BHZ Hamiltonian}) by replacing the continuous momentum operators $\hat{k}_{x,y}\Psi(x,y)=-i \partial_{x,y}\Psi(x,y)$ by their discretized versions in the Hamiltonian of Eq.~(\ref{BdG BHZ Hamiltonian})\cite{Datta1995a}
\begin{align}
\label{DerivativeDis}
    \left.\partial_{x/y}\Psi\right|_{j_{x/y}}&\approx \frac{1}{2a_{x/y}}\left( \Psi_{j_{x/y}+a_{x/y}}-\Psi_{j_{x/y}-a_{x/y}} \right),\\
\label{DoubleDerivativeDis}
    \left.\partial^2_{x/y}\Psi\right|_{j_{x/y}}&\approx \frac{1}{a^2_{x/y}}\left( \Psi_{j_{x/y}+a_{x/y}}-2\Psi_{j_{x/y}}+\Psi_{j_{x/y}-a_{x/y}} \right),
\end{align}
where $a_{x/y}=a=5\,$nm are the lattice constants of the two-dimensional lattice and $j_{x/y}$ is the site index in $x/y$ direction. 
\subsection{Transport formalism}
We introduce the formalism we use to calculate the conductance, the spectral density and the critical current for the NS and SNS junction, respectively. The differential conductance at voltage $V$ is given by 
\begin{equation}
    \label{Zero-bias conductance}
    G(V) = G_0 \int_{-\infty}^\infty d\varepsilon \left(-\frac{\partial f(\varepsilon-eV)}{\partial \varepsilon} \right) (N_{\text{tot}}-{\cal R}^{ee}+{\cal R}^{he}),
\end{equation}
with  $N_\text{tot}$ the total number of modes, $G_0 = e^2/h$, $f(\varepsilon)$ the Fermi-Dirac distribution, and the Andreev reflection probability 
\begin{equation}
    \label{eq:Caroli}
    {\cal R}^{\alpha \beta}(\varepsilon)=\mathrm{Tr} \left[
    \Gamma_{N,\alpha} G^{a}_{\alpha \beta} \Gamma_{N,\beta} G^{r}_{\beta \alpha}
    \right].
\end{equation}
with the retarded/advanced GFs
\begin{align}
    G^{r/a}=\left(\varepsilon\pm i \eta-H_C-\Sigma^{r/a}_N-\Sigma_S^{r/a}\right)^{-1}.
\end{align}

Here, $H_C$ is the Hamiltonian of a slice of the two-dimensional system placed on the normal part and $V_{C,\alpha}=V_{\alpha,C}^\dagger$ are the coupling matrices between the central and the $\alpha-$parts. To calculate $G^{r/a}$, we use standard recursive Green's functions techniques~\cite{Lopez1984a, MacKinnon1985, Dominguez2024Fraunhofer, Traverso2026a}. 
Moreover, we have introduced the retarded/advanced self-energies $\Sigma_{\alpha}^{r/a}(\varepsilon) = V_{C,\alpha} g_{\alpha}^{r/a} (\varepsilon )V_{\alpha,C}$ of the $\alpha=N,S$ lead and their corresponding rates $\Gamma_{\alpha} =i\left(\Sigma_\alpha^r-\Sigma_\alpha^a\right)$. If not stated otherwise, we perform all calculations in the zero temperature limit.

For the SNS junction, we focus on the integrated spectral density over the selected normal region at energy $E$ from which we extract the Andreev bound state spectrum. We do this by taking the imaginary part of the retarded Green's function as
\begin{align}
    \rho_{n}(E)=-\frac{1}{\pi} \text{Im}\sum_{x=1}^{N}\text{Tr}_{W}\{G^{r}(x,x,E)\},
\end{align}
with $N$ as the number of sites in the central part and $\text{Tr}_{W}$ as the trace over the width $W$. 

Within the same framework, we evaluate the equilibrium supercurrent as~\cite{martinrodero1994, Yeyati1995a}
\begin{align}
\label{eq:statcurr}
I(\varphi) &= \frac{e}{\hbar} \int dE~ \text{Tr}_W \Big\{ 
[V_{LR} G^{+-}_{RL}(E) \nonumber \\
&\qquad\quad - V_{RL} G^{+-}_{LR}(E)]_e \Big\}, 
\end{align}
where $G_{RL}^{+-}(E)\equiv G^{+-}(x-a,x+a,E,V=0)$ are the equilibrium lesser Green's function at $L=x-a$ and $R=x+a$ and the subindex $e$ represents the electronic part of the system.
From Eq.~(\ref{eq:statcurr}), we obtain the critical current as the maximal equilibrium current for a given superconducting phase difference $\varphi$
\begin{align}
    I_c&=\text{Max}_{\varphi}\ | I(\varphi)|,
    \end{align}
where $I(\varphi,V=0)$ is the supercurrent of the junction. 
\section{NS junction}
\label{NS junction}

In this section, we analyze the normal-backscattering mechanism generated by a potential barrier adjacent to the superconducting contact. We characterize this mechanism through the dependence of the zero-bias conductance on the different model parameters. Depending on the strength of the disorder potential in the barrier, we distinguish between ballistic and disordered regimes. 

\begin{figure}[tb]
  \centering
  \includegraphics[width=3.3 in]{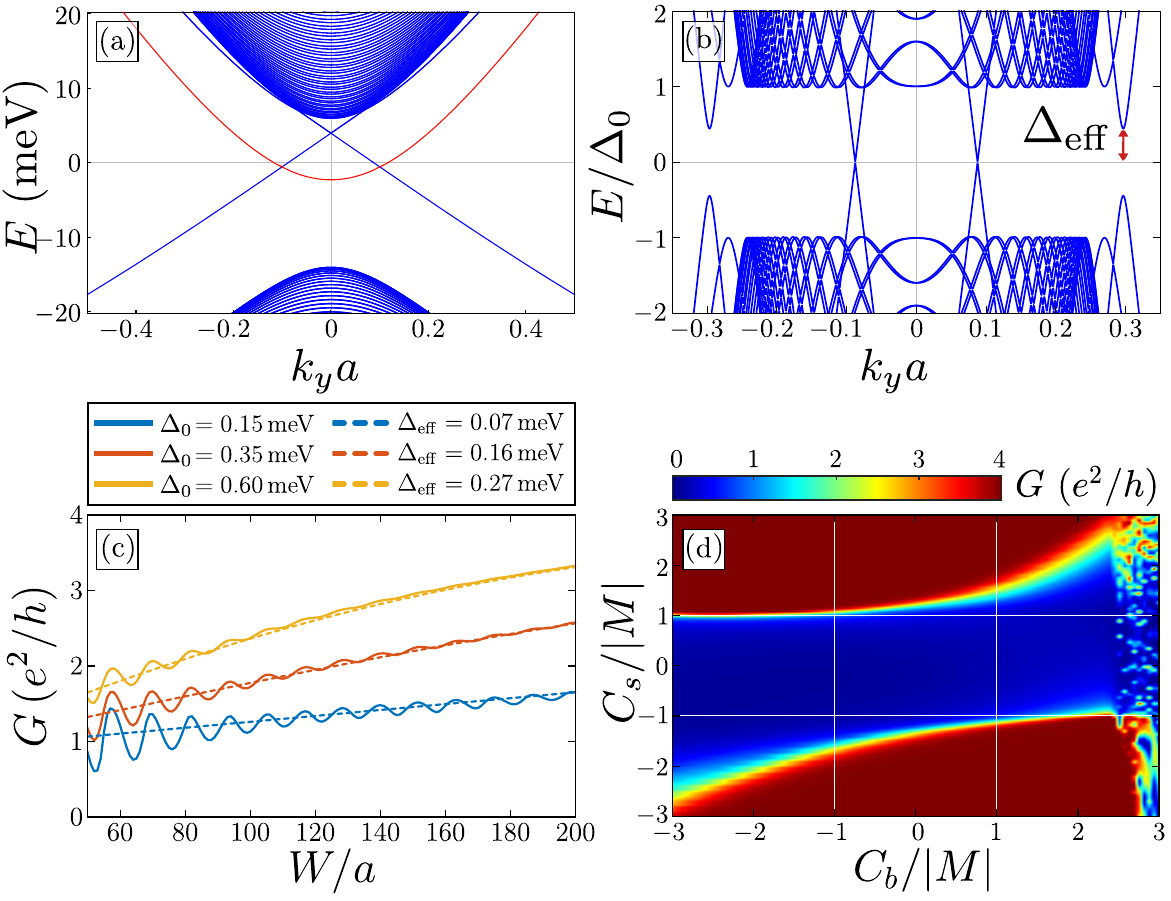}
  \caption{(a) Energy spectrum $E(k_y)$ of a QSH/barrier/QSH junction with $C_n=-4\,\mathrm{meV}$, $C_b=-20\,\mathrm{meV}$, and $L_b/a=11$. (b) Energy spectrum $E(k_y)$ of a QSH/barrier/SC junction with $C_n=-4\,\mathrm{meV}$, $C_b=19\,\mathrm{meV}$, $C_s=20\,\mathrm{meV}$, $L_b/a=5$, and $M_s=10\,\mathrm{meV}$. (c) Conductance $G$ as a function of the junction width $W$ for different values of the superconducting pairing amplitude $\Delta_0$. Solid lines show the numerical results, while dashed lines indicate fits based on Eq.~(\ref{Normal reflection}). (d) Conductance $G$ as a function of the chemical potential $C_s$ in the superconductor and the barrier chemical potential $C_b$, with $\Delta_0=0.15\,$meV, $W/a=200$ and $L_b/a=5$. The white lines represent the onset of the bulk bands.}
  \label{CouplingEdgesMap}
\end{figure}

\subsection{Ballistic regime}

To identify the conditions under which the top and bottom QSH edges can couple, we begin with a reference (non-superconducting) QSH/barrier/QSH junction. The barrier is also described by the BHZ model, but its chemical potential lies within either the conduction or valence bulk band, controlled by $C_b$. In this regime, a reduced number of bulk bands can connect the two edges, see the red band in Fig.~\ref{CouplingEdgesMap}(a). Because these states extend across the entire barrier width, they mediate coherent coupling between opposite edges. In this fully normal junction, repeated propagation and reflection at the two QSH/barrier interfaces produce Fabry--P\'{e}rot-type interference. Consequently, the interedge coupling, and hence the conductance, oscillates as a function of both $W$ and $L_b$. The width of the corresponding resonances is controlled by the transparency of the QSH/barrier interfaces, see further details in App.~\ref{QSHI/barrier/QSHI-junction}.

The backscattering mechanism changes qualitatively when the right-hand QSH region is replaced by a superconductor, forming a QSH/barrier/SC junction. The barrier bands that couple to the superconducting region become partially proximitized and acquire an induced gap $\Delta_{\mathrm{eff}}\leq\Delta_0$, as illustrated in Fig.~\ref{CouplingEdgesMap}(b). Low-energy propagation across the barrier then becomes evanescent, acquiring an exponentially decaying envelope, $\mathcal{A}_{\mathrm{tb}}\sim e^{-W/\xi_{\mathrm{eff}}}$,
where
\begin{align}
    \xi_{\mathrm{eff}}
    =\frac{\hbar v_{F,\mathrm{barrier}}}{\Delta_{\mathrm{eff}}}
\end{align}
is the effective superconducting coherence length and
$v_{F,\mathrm{barrier}}$ is the Fermi velocity of the relevant barrier mode. 

Coupling the top and bottom QSH edges opens a normal-backscattering channel and reduces the zero-bias conductance below its quantized value, $G=4e^2/h$. Increasing the barrier width $W$ suppresses this coupling and drives the conductance toward $4e^2/h$, as shown in Fig.~\ref{CouplingEdgesMap}(c). Increasing the parent pairing amplitude $\Delta_0$ has a similar effect: the induced gap $\Delta_{\mathrm{eff}}$ becomes larger, thereby reducing $\xi_{\mathrm{eff}}$ and weakening the coupling between opposite edges. Additionally, finite-size interference produces the residual oscillations visible in Fig.~\ref{CouplingEdgesMap}(c), but the overall decay of the interedge coupling is controlled by $\xi_{\mathrm{eff}}$. Since $\Delta_{\mathrm{eff}}$ depends on both the barrier geometry and the transparency of the barrier--superconductor interface, it can be considerably smaller than the parent gap $\Delta_0$. The resulting $\xi_{\mathrm{eff}}$ may therefore become comparable to the barrier widths of typical experimental devices.

 To gain insight into the width dependence, and to stay within the edge state picture, we represent the hybrid interface phenomenologically by an equivalent single-channel normal scatterer, with transmission (reflection) probability $T_{\mathrm{eff}}=1-R_{\mathrm{eff}}$, followed by an ideal Andreev reflector. At zero energy, the corresponding Andreev reflection probability reads
 \begin{align}\label{Andreev conductance}
    \mathcal{R}_{he}^{\mathrm{NS}}
    =\left(\frac{T_\text{eff}}{2-T_\text{eff}} \right)^2=
      \left(\frac{1-R_{\mathrm{eff}}}
           {1+R_{\mathrm{eff}}}\right)^2,
 \end{align}
with $G=(4e^2/h)\mathcal{R}_{he}^{\mathrm{NS}}$. Here, the factor $T_{\mathrm{eff}}^2$ reflects the fact that both the incident electron and the Andreev-reflected hole must traverse the effective scatterer, while the denominator accounts for repeated normal reflections\cite{Nazarov2009}. Note that $R_{\mathrm{eff}}$ should not be confused with the electron-to-electron reflection probability $\mathcal{R}_{ee}^{\mathrm{NS}}$ of the complete superconducting junction. At zero energy and in the single channel case, unitarity relates both by Eq.~\eqref{Andreev conductance} and $\mathcal{R}_{ee}^{\mathrm{NS}}+\mathcal{R}_{he}^{\mathrm{NS}}=1$.

The equivalent scatterer parametrizes the nonlocal
interedge process mediated by the proximity-gapped
barrier modes. Assuming that the reflection amplitude of the equivalent scatterer inherits the evanescent envelope of the microscopic interedge propagation amplitude $\sim\exp(-W/\xi_{\mathrm{eff}})$, we write 
\begin{align}
    \label{Normal reflection}
    R_{\mathrm{eff}}
    =\rho_0 e^{-2W/\xi_{\mathrm{eff}}}.
\end{align}
The prefactor $\rho_0$ contains the remaining microscopic information about the geometry and the effective coupling among the QSH edges, the barrier, and the superconducting region. We use Eqs.~\eqref{Andreev conductance} and~\eqref{Normal reflection} to fit the conductance as a function of $W$ for different values of $\Delta_0$. 
For each value of $\Delta_0$, the induced gap $\Delta_{\mathrm{eff}}$ and the velocity $v_{F,\mathrm{barrier}}$ are extracted from the corresponding band structure, as illustrated in Fig.~\ref{CouplingEdgesMap}(b), and used to determine $\xi_{\mathrm{eff}}$. A common value of $\rho_0$ is sufficient to describe all three curves, indicating that the dominant dependence on the parent pairing amplitude is captured by the corresponding variation of $\xi_\text{eff}$. The resulting fits, shown by the dashed lines in Fig.~\ref{CouplingEdgesMap}(c), reproduce the overall width dependence of the conductance, while the numerical results exhibit additional finite-size oscillations. 

\begin{figure}[tb]
  \centering
  \includegraphics[width=3.3 in]{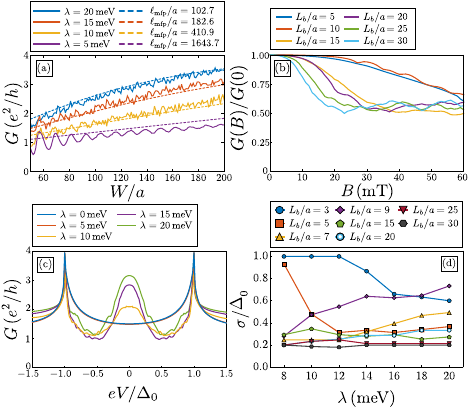}
  \caption{(a) Zero-bias conductance $G$ versus junction width $W$ for several
disorder strengths $\lambda$. Solid lines are numerical results, dashed lines
fits to Eqs.~\eqref{Andreev conductance}-\eqref{Diffusion length}. (b) Conductance normalized to its
zero-field value, $G(B)/G(0)$, versus magnetic field $B$ for several barrier
lengths $L_b$. (c) $G$ versus bias voltage $eV$ for several $\lambda$. (d) Half width at half maximum
$\sigma$ of the zero-bias peak in (c) versus $\lambda$, for $L_b/a=3\ldots 30$.
Unless stated otherwise, $L_b/a=5$, $W/a=150$, $eV=0$, $B=0$, $\Delta_0=0.15\,$meV. All curves are averaged over $100$ disorder realizations, with
the same set of realizations used for every parameter value.
}
  \label{DisorderConductance}
\end{figure}

Finally, Fig.~\ref{CouplingEdgesMap}(d) shows the conductance as a function of the chemical potentials in the barrier and superconducting regions, $C_b$ and $C_s$, respectively, for $\Delta_0=0.15\,\mathrm{meV}$ and $L_b=5a$. A zero-conductance region appears for $|C_s/M|<1$, where the superconducting region is in its insulating regime. Outside this gap, and particularly for $|C_s/M|\gg1$, the conductance approaches its quantized value, $G=4e^2/h$.

Close to the band edges, $|C_s/M|\simeq1$, the conductance depends strongly on the relative signs of $C_b$ and $C_s$. For opposite signs, the conductance rapidly approaches $4e^2/h$, whereas for equal signs it remains suppressed over a considerably broader parameter range. This asymmetry originates from the different $E_1$- and $H_1$-orbital character of the conduction and valence bands. Because $M_s>0$ in the superconducting region while $M_b<0$ in the barrier, bands with the same electron- or hole-like doping have opposite orbital character in the two regions. The resulting orbital mismatch weakens the coupling across the barrier--superconductor interface and reduces $\Delta_{\mathrm{eff}}$, $T_\text{eff}$, and consequently the conductance.

\subsection{Disordered regime}

Having established the backscattering mechanism in the ballistic limit, we now examine how it is modified by disorder within the barrier, see Fig.~\ref{DisorderConductance}(a). Here, we can observe an enhancement of the conductance and its slope as a function of $W$ for an increasing disorder strength $\lambda$. Although an increase of conductance with disorder may appear counterintuitive, it follows from the fact that at zero energy and in the presence of time-reversal symmetry, an Andreev-reflected hole follows the time-reversed trajectory of the incoming electron. The phases accumulated along the electronic and hole trajectories then cancel, allowing the corresponding scattering sequences to interfere constructively. Disorder can therefore suppress interedge backscattering while simultaneously enhancing local Andreev reflection. This mechanism is analogous to \textit{reflectionless tunneling}, originally
observed as a zero-bias conductance peak in superconductor--semiconductor
junctions~\cite{Kastalsky1991,Nguyen1992,Bakker1994,Magnee1994} and subsequently
analyzed both numerically and analytically~\cite{vanWees1992,Takane1992,
Marmorkos1993,Takane1993,Volkov1993,Hekking1993,Beenakker1994,Nazarov1994}.

The effective description of the NS conductance in Eqs.~\eqref{Andreev conductance}--\eqref{Normal reflection} is modified in the presence of disorder. Here, an additional length scale enters: the mean free path $\ell_{\mathrm{mfp}}$. In this scenario, the ballistic $\xi_{\mathrm{eff}}$ becomes attenuated by $\ell_{\mathrm{mfp}}$, that is,
\begin{align}
\label{Diffusion length}
\frac{1}{\xi_{\mathrm{att}}}
=
\frac{1}{\xi_{\mathrm{eff}}}
+\frac{1}{2\ell_{\mathrm{mfp}}}.
\end{align}
Since $\xi_{\mathrm{att}}<\xi_{\mathrm{eff}}$, the conductance is enhanced within this effective description. Using the same fitting parameters as in the ballistic regime, together with the estimated values of $\ell_{\mathrm{mfp}}$, we obtain the dashed curves shown in Fig.~\ref{DisorderConductance}(a). Further details are provided in App.~\ref{app.effectivediff}.

\begin{figure*}[tb]
  \includegraphics[width=1\textwidth]{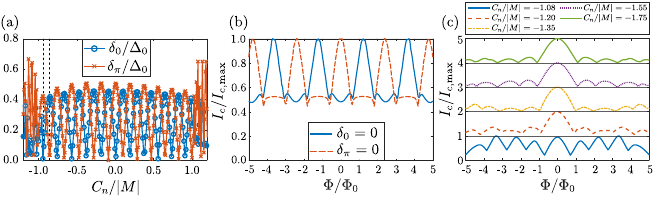}
  \caption{SNS junction:
(a) Gap openings at $\varphi=0$ and $\varphi=\pi$, denoted by $\delta_0$ and $\delta_\pi$, as a function of the chemical potential $C_n$ for $L_b=5a$.
(b) Corresponding SQI patterns evaluated at the values of $C_n$ indicated by the dashed lines in panel (a).
(c) SQI patterns for values of $C_n$ in the conduction band, again for $L_b=5a$. In this panel, all curves are normalized to unity and vertically offset for clarity.
  }
  \label{ABSSQIFraunhofer}
\end{figure*}

The conductance enhancement described above relies on the constructive interference between electron and hole paths.  
Hence, perturbations introducing a relative phase shift between the electron and the retroreflected hole can suppress the enhancement originated by the presence of disorder. Here, we consider two such perturbations: a perpendicular magnetic field and a finite bias voltage.

A perpendicular magnetic field, $B$, introduces a relative Peierls phase between the electronic and hole path. To model this, we incorporate the magnetic field through the minimal substitution $ \hat{k}_{x,y}\longrightarrow \hat{k}_{x,y}-\frac{e}{\hbar}A_{x,y}$, with the Landau gauge
$\boldsymbol{A}=-B y\,\boldsymbol{e}_x$ at the normal part of the junction. In Fig.~\ref{DisorderConductance}(b), we show the normalized conductance $G(B)/G(0)$ as a function of the applied magnetic field. As we have anticipated, we observe a reduction of the conductance as a function of $B$. Eventually, this reduction saturates for magnetic fields above some critical value $B>B_c$, determined by the geometry of the barrier, $B_c\sim \Phi_0/L_bW$, with $\Phi_0=h/2e$. 

A finite bias voltage likewise introduces a relative dynamical phase between the electron and hole trajectories, suppressing constructive interference. As a result, the conductance develops a zero-bias peak (ZBP), see Fig.~\ref{DisorderConductance}(c). In Fig.~\ref{DisorderConductance}(d), we extract the half width at half maximum $\sigma$ of the ZBP as a function of $\lambda$ and for different $L_b$. We find a reduction of $\sigma$ for increasing $\lambda$ when the barrier contains one proximitized band $L_b\lesssim5a$. In contrast, for $L_b>5a$, we observe a reduced and rather constant dependence of $\sigma$ on $\lambda$. Previous analysis of the ZBP relates its width to the dwell time that a quasiparticle spends in the disordered region, i.e.~$\sigma\sim eV_c= \hbar/\tau_{\text{dwell}}$. Here, $\tau_{\text{dwell}}$ depends on the mean level spacing $\delta=\left[d(E_F)L_bW\right]^{-1}$, with the number of escape channels and the effective transparency in the absence of the reflectionless-tunneling enhancement and $d(E_F)$ density of states at the Fermi level, see further details in App.~\ref{App:Long barrier}. In the present system, this estimation is most reliable when several barrier bands are present at the Fermi energy. In the short barrier regime, $L_b/a<10$, only a single band connects opposite edges, making the width of the ZBP depend non-monotonically on $L_b$ and $\lambda$.

\section{SNS junction} \label{Andreev bound states and SQI pattern} 

As discussed in the previous section, coupling top and bottom QSH edges enables quasiparticles to backscatter via the opposite edge. In a Josephson junction with two such NS interfaces, this backscattering lifts the protection of the crossings at $\varphi=0$ and $\varphi=\pi$, opening the corresponding gaps $\delta_0$ and $\delta_\pi$, see Fig.~\ref{Fig1}(d).

The amplitude of these gaps depends not only on how strongly the edges are coupled, but also on the phases acquired while propagating through the normal part of the junction. As a result, $\delta_0$ and $\delta_\pi$ oscillate as a function of $C_n$ and $L_n$, in close analogy with a particle traversing a double-barrier potential. An example is shown in Fig.~\ref{ABSSQIFraunhofer}(a), where we extract $\delta_0$ and $\delta_\pi$ from the ABS spectrum as a function of $C_n$. Here, we observe an oscillatory pattern for both $\delta_0$ and $\delta_\pi$, with a relative phase shift such that one gap is maximal when the other is minimal. The period of these oscillations is set by the phase acquired while propagating through the normal part. In the limit where the dynamical phases are negligible, this phase is given by 
\begin{align}
\label{Resonance condition}
\Psi= 2(C_n/E_{T,n}+C_b/E_{T,W}) ,
\end{align}
where $E_{T,n}=\hbar v_F/L_n$ and $E_{T,W}=\hbar v_{F,\text{barrier}}/(W+2L_b)$ are the Thouless energies, $v_{F,i}$ is the Fermi velocity at the $i$th-sector. Thus, the Fabry-P\'{e}rot resonance condition is achieved when $\Psi=2\pi n$, with $n\in \mathbb{Z}$. 

Using a simplified version of the scattering network model presented in Ref.~\onlinecite{Baxevanis2015a}, we can analytically relate $\delta_\pi$ to the transmission coefficient $T$ of the normal junction through the energy-phase relation
\begin{align}
E_\pm(\varphi)=\pm\Delta_0 \sqrt{1-T\sin^2(\varphi/2)} ,
\end{align}
with the conventional double barrier transmission function~\cite{Datta1995a} 

\begin{align}
T=\frac{T_{\text{eff},L} T_{\text{eff},R}}{1+R_{\text{eff},L}R_{\text{eff},R}-2\sqrt{R_{\text{eff},L}R_{\text{eff},R}}\cos(\Psi)},
\label{eq.transmissionSNS}
\end{align}
with $T_{\text{eff},L/R}=1-R_{\text{eff},L/R}$ the effective transmission probabilities at the left/right NS interfaces and $\Psi$ is given in Eq.~\eqref{Resonance condition}. Although $\Psi$ is generally energy dependent, we approximate it here as energy independent.

In the short-junction limit, $\delta_0=0$  and $\delta_\pi=E_+(\pi)-E_-(\pi)=2\Delta_0\sqrt{1-T}$. However, when additional ABSs are present within the gap, $\delta_0$ becomes finite, with a maximum value of the order of $\delta_\pi$, with the mentioned phase shifted relative to $\delta_\pi$. We use the scattering network model to capture the intermediate and long-junction regimes, see further details in App.~\ref{Scattering network calculations}.

Coupling opposite edges modifies the trajectories of the ABSs and, consequently, the areas they enclose. This directly affects the SQI pattern, namely, the critical current as a function of an applied perpendicular magnetic field~\cite{Baxevanis2015a,Haidekker2020a,Vigliotti_2022,Viglioti2023a}. In this scenario, the initial SQUID-like pattern with flux period $\Phi_0$, obtained for decoupled edges, evolves into a pattern with a doubled flux period, $2\Phi_0$, when the edges are coupled. This so-called even--odd effect manifests itself as a suppression of every other SQUID lobe, as shown in Fig.~\ref{ABSSQIFraunhofer}(b). 
We find the relation of the ratio $\delta_0/\delta_\pi$ in connection to the suppression of the lobes. Indeed, the size of this ratio correlates with the parity of the suppressed lobe: for $\delta_0/\delta_\pi<1$ ($\delta_0/\delta_\pi>1$), lobes centered at $\Phi/\Phi_0=2 n$ ($\Phi/\Phi_0=2 n+1$), with $n\in \mathbb{Z}$, become suppressed. 
Remarkably, in the presence of reflection symmetry $(T_{\text{eff},L}=T_{\text{eff},R})$, we can tune $\Psi$, such that $\delta_\pi=0$ while decoupling the ABS from the quasicontinuum ($\delta_0\neq0$). Under these and parity conserving conditions, the measurement of a fractional Josephson effect is possible. Indeed, one can track this sweet-spot, by measuring a SQI pattern when the lobes placed at $\Phi/\Phi_0=2 n+1$ become strongly suppressed, see Fig.~\ref{ABSSQIFraunhofer}(b). 

The SQUID-like pattern turns progressively into a Fraunhofer-like pattern when setting the chemical potential within the first bulk bands of the QSH part, see Fig.~\ref{ABSSQIFraunhofer}(c). For $|C_n/M|\gtrsim 1$, we can observe that some of the Fraunhofer lobes inherit the even-odd effect of the SQUID pattern. Besides, the interplay of the Fraunhofer pattern and the SQUID pattern gives rise to a non-monotonic decay and local minima within the Fraunhofer lobes, see Fig.~\ref{ABSSQIFraunhofer} (c). Despite the Fraunhofer-like character of the SQI pattern, we can observe that the even–odd lobe modulation is still visible for $|C_n/M|\gtrsim 1$ and also that the lobes exhibit local minima~\cite{Pribiag2015a}.

\begin{figure}[t!] 
\centering \includegraphics[width=1\linewidth]{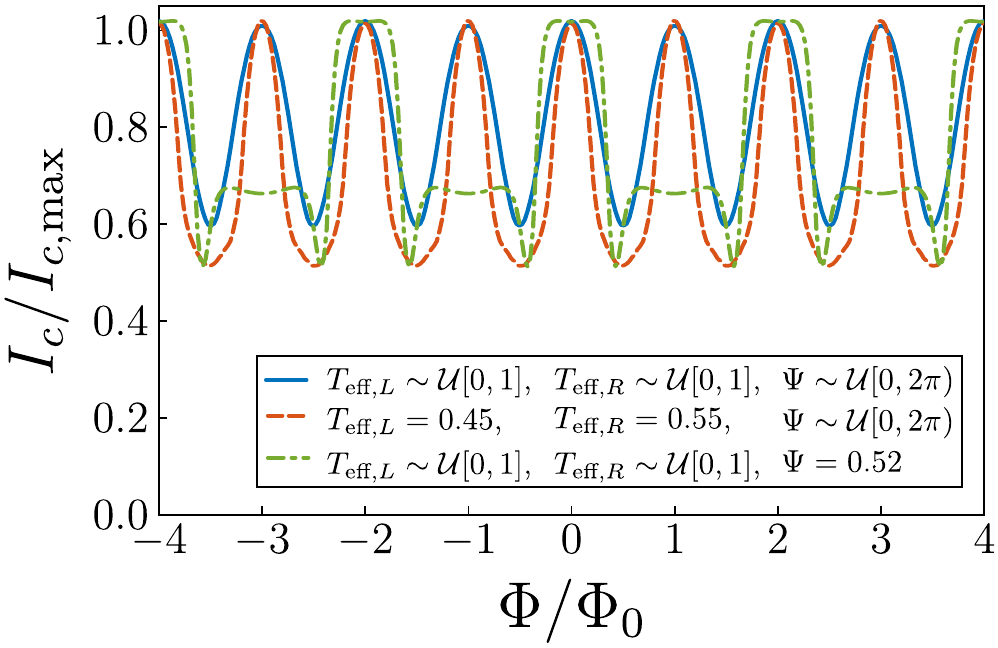} 
\caption{Disorder-averaged critical current $I_c$ versus the magnetic flux $\Phi$ obtained from the scattering network model. The solid blue curve corresponds to independently randomized effective transmissions $T_{\text{eff},L/R}$ and phase $\Psi$. For the dashed red line $T_{\text{eff},L/R}$ are fixed and only $\Psi$ is randomized. For the green dashed line $T_{\text{eff},L/R}$ are randomized and $\Psi$ is fixed.
Here, $x \sim \mathcal{U}[a,b)$ means that $x$ is drawn uniformly at random from the interval $[a,b)$
Each curve is averaged over 1000 disorder configurations. } \label{SQIDisorder} \end{figure}

Potential disorder in the normal regions randomizes the phase $\Psi$ and the effective transmission probabilities $T_{\text{eff},L/R}$. Despite the finite coupling, the combination of both effects contributes to restoring the $\Phi_0$-periodicity of the SQI pattern. To see this effect, we use the phenomenological scattering network model and obtain the SQI for disorder realizations on $\Psi$, see Fig.~\ref{SQIDisorder} and App.~\ref{Results of the scattering network} for more details. Thus, our results show that the absence of an even-odd lobe modulation should not be taken as evidence for decoupled edges.

\section{Conclusions}
\label{sec:conclusions}

We have studied the coupling between opposite helical edges in quantum spin Hall NS and SNS junctions containing a conducting region adjacent to the superconducting contacts. Within a microscopic BHZ model in the Bogoliubov--de Gennes formalism, this region provides a time-reversal-symmetric backscattering mechanism: a quasiparticle can reverse its propagation direction by transferring to the opposite edge, without requiring local backscattering within an individual helical channel. The strength of this process is determined by the properties of the complete hybrid interface and therefore cannot, in general, be inferred from the reflection probability of the corresponding isolated normal barrier.

For an NS junction, interedge backscattering reduces the zero-bias Andreev conductance below the quantized value $G=4e^2/h$. When the conducting barrier is coupled to the superconductor, its transverse modes acquire an induced gap and mediate an evanescent coupling between opposite edges. The resulting width dependence is controlled by an effective coherence length set by the induced gap and the velocity of the relevant barrier mode. An effective-transparency description reproduces the overall decay of the interedge contribution, while the microscopic calculations retain finite-size oscillations. We further find that the conductance depends on the barrier length and doping, the coupling to the superconducting region, and the orbital and Fermi-velocity mismatch across the interface. In the presence of disorder, the zero-bias conductance is enhanced within the parameter range considered and is reduced by a finite bias voltage or perpendicular magnetic field. This behavior is consistent with coherent electron--hole interference of the type underlying reflectionless tunneling.

We have subsequently analyzed how the same interedge process modifies an SNS Josephson junction. Coupling states associated with opposite edges generically opens gaps between the relevant Andreev levels at the time-reversal-invariant phase differences $\varphi=0$ and $\varphi=\pi$. The corresponding gaps, denoted by $\delta_0$ and $\delta_\pi$, oscillate with the chemical potential and junction geometry because of the dynamical phase accumulated during propagation through the normal and barrier regions. Since these oscillations are shifted relative to one another, one of the gaps can close while the other remains finite. A closing at a single phase difference therefore does not, by itself, imply that the two edges are decoupled. The scattering-network description reproduces this behavior and relates the gap oscillations to the transmission resonances of the normal junction.

The same phase dependence is reflected in the superconducting quantum interference pattern. In the edge-dominated regime, interedge scattering produces an even--odd modulation of the SQUID-like lobes, with the parity of the more strongly suppressed lobes correlated with the relative sizes of $\delta_0$ and $\delta_\pi$. When the chemical potential enters the bulk bands, the interference pattern evolves toward a Fraunhofer-like response, while residual edge contributions can remain visible through local minima and an unequal modulation of neighboring lobes.

Taken together, these results connect the microscopic properties of the superconducting interface to the NS conductance, the Andreev spectrum, and the magnetic interference pattern. These observables provide complementary information about interedge scattering: the NS conductance primarily characterizes its magnitude, whereas the Josephson spectrum and SQI pattern are also sensitive to the associated scattering phase. This comparison provides a practical basis for assessing departures from the independent-edge description in quantum spin Hall--superconductor junctions.

\begin{acknowledgments}
We acknowledge stimulating discussions with M.~P. Stehno, E.~M. Hankiewicz, B. Trauzettel, C. Gould, H. Buhmann, and L.~W. Molenkamp. FD acknowledges funding support from the Deutsche Forschungsgemeinschaft (DFG, German Research Foundation) under Germany’s Excellence Strategy through the W\"urzburg-Dresden Cluster of Excellence ctd.qmat (EXC 2147, Project ID 390858490) as well as through the Collaborative Research
Center SFB 1170 ToCoTronics (Project ID 258499086).
\end{acknowledgments}

\bibliography{bibliography.bib}

\pagebreak

\appendix

\section{QSH/barrier/QSH-junction}
\label{QSHI/barrier/QSHI-junction}

To determine when the top and bottom edges of a QSH can couple, we first study a QSH/barrier/QSH-junction~\cite{Reinthaler2013a}. We model the barrier as a QSH region with a doping $C_b$ different from that of the outer QSH regions $C_n$. We keep the chemical potential of the outer QSH edges inside the bulk gap by setting $C_n=-4\,$meV, while varying the barrier chemical potential $C_b$. 

\begin{figure}[t!]
  \centering
  \includegraphics[width=1\linewidth]{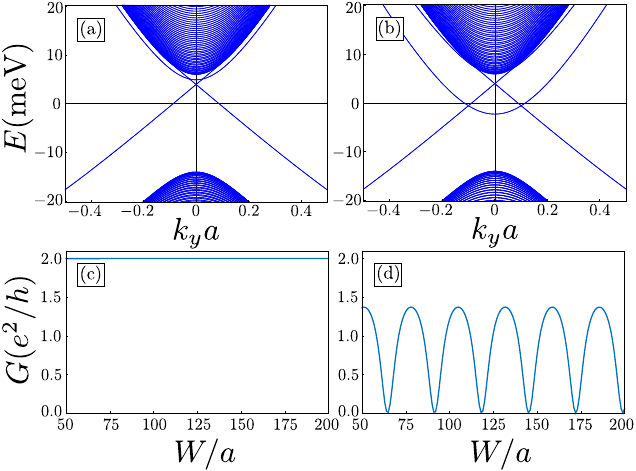}
  \caption{Top row: the dispersion relation for a QSH/barrier/QSH junction, with $C_n=-4\,$meV and (a) $C_b=-8\,$meV and (b) $C_b=-20\,$meV. Bottom row: the corresponding zero-bias conductance $G$ as a function of the width $W$ of the system.
  }
  \label{QSHIbarrierQSHI}
\end{figure}

The top and bottom QSH edges are decoupled, as long as the barrier does not exhibit bulk states at the Fermi energy, as shown in Fig.~\ref{QSHIbarrierQSHI}(a). 
As a consequence, the conductance stays quantized to the value of $G=2e^2/h$ for all values of the width, as illustrated in Fig.~\ref{QSHIbarrierQSHI}(c). Then, setting the chemical potential of the barrier such that a bulk band is at the Fermi energy, the top and bottom edges become coupled through the extended bulk states, see Fig.~\ref{QSHIbarrierQSHI}(b). In this situation, the corresponding zero-bias conductance deviates from the quantized value of $2e^2/h$ and exhibits oscillations with the system width, as shown in Fig.~\ref{QSHIbarrierQSHI}(d). 

To get a better understanding of the scaling of the edge coupling mechanism, we describe the interedge scattering as a series of coherent propagation and reflection events inside the barrier. The amplitude for a particle incident on the top edge to scatter into the bottom edge is
\begin{align}
\nonumber
    A_{T\rightarrow B}&=t_T t_B e^{ik_F^bW}\sum_{n=0}(r_Tr_Be^{2ik_F^bW})^n\\
    \label{Transmission amplitude}
    &=\frac{t^2 e^{i(k_F^bW+\chi)}}{1-r^2be^{i(2k_F^bW+\phi)}} 
\end{align}
where $t_T$ and $t_B$ describe the amplitudes of a particle entering the barrier at the top edge and leaving it at the bottom edge, respectively. Similarly, $r_T$ and $r_B$ represent the amplitudes for a particle reflecting at the top or bottom of the barrier, respectively. Every time a particle traverses the full length of the barrier, it picks up a phase factor of $e^{ik_{F}^bW}$, where $k_F^b$ is the Fermi wave vector of the bulk band. In the last line, we assumed for simplicity that the amplitudes have the same absolute value but can differ in a phase $t_tt_b=t^2e^{i\chi}$ and $r_tr_b=r^2 e^{i\phi}$.

Due to the presence of time reversal symmetry, there will be an equal probability amplitude that couples bottom to top, i.e.,~$A_{B\rightarrow T}$. This will be accounted for by a factor 2 in the conductance of the system.

With these scattering amplitudes we compute the backscattering probability  as 
\begin{align}
    R\equiv|A_{T\rightarrow B}|^2=\frac{(1-|r|^2)^2}{1+|r|^4-2|r|^2\cos\left(2k_{F}^bW+\phi\right)},
\end{align}
Finally, we can express the conductance at zero bias voltage of the QSH/barrier/QSH junction as
\begin{align}
\nonumber
    G&=\frac{2e^2}{h}(1-R)\\
    \label{Fabry-Perot-Conductance}
    &=\frac{2e^2}{h}\frac{F \sin^2(k_F^bW+\phi/2)}{1+F \sin^2(k_F^bW+\phi/2)}
\end{align}
with $F=4|r|^2/(1-|r|^2)^2$ as the coefficient of finesse, in analogy with the standard Fabry-P\'{e}rot finesse. By fitting $r$ and $\phi$ in Eq.~(\ref{Fabry-Perot-Conductance}), we reproduce the numerical results shown in Figs.~\ref{QSHIbarrierQSHI} (c) and (d). 

\section{Effects of spin non-conserving SOC and group velocity mismatch}
\label{SOC and Fermi velocity mismatch}

In this section, we discuss further details that can modify the backscattering probability at the NS junction: the presence of spin non-conserving scattering by the addition of Dresselhaus SOC and the group velocity mismatch between the superconductor and the barrier. 
We start with the effect of spin-orbit coupling on the conductance by introducing Dresselhaus SOC of the form 
\begin{align}
    \nonumber
    \mathcal{H}_D&=\delta_{\text{SOC}} \sigma_y s_y-\delta_e \frac{\sigma_0+\sigma_z}{2}(\hat{k}_xs_x-\hat{k}_y s_y)\\
    &+\delta_h \frac{\sigma_0-\sigma_z}{2}(\hat{k}_xs_x+\hat{k}_ys_y)
\end{align}
where $\delta_{\text{SOC}}=1.6\,$meV, $\delta_e=-12.8\,$meV nm and $\delta_h=21.1\,$meV\,nm are the bulk inversion asymmetry parameters~\cite{Pikulin2014}. 

\begin{figure}[t!] \centering \includegraphics[width=1\linewidth, height=0.45\linewidth]{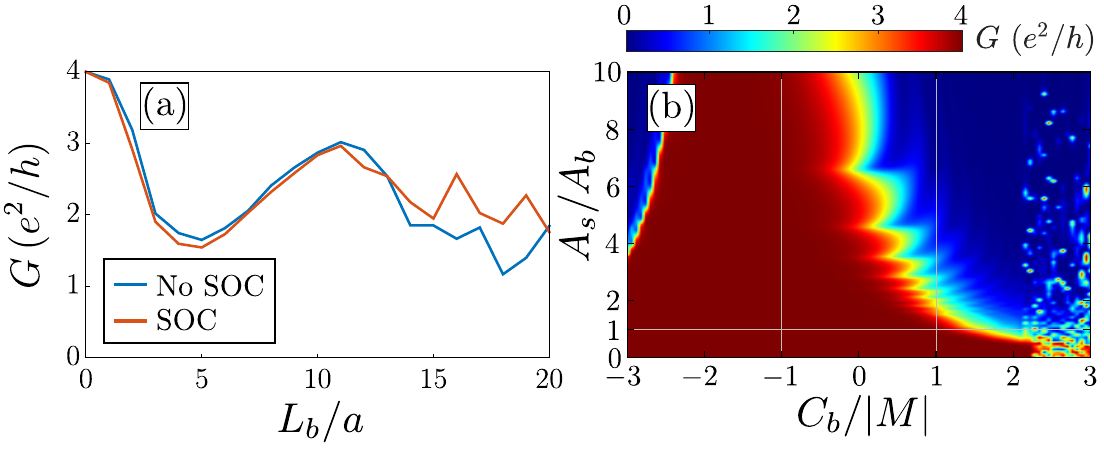} \caption{(a) Normal conductance $G$ as a function of the barrier length $L_b$ with and without spin-orbit coupling. (b) Normal conductance $G$ as a function of the barrier chemical potential $C_b$ and the ratio $A_s/A_b$, where $A_s$ and $A_b$ denote the $A$ parameters in the superconducting region and in the barrier, respectively.} 
\label{SOCandvFMismatch}
\end{figure}

We consider the presence of SOC only the QSH and the barrier and observe only a small change in the conductance for $L_b/a\leq10$, where only one band is present at the Fermi energy, as illustrated in Fig.~\ref{SOCandvFMismatch} (a). When the second band enters in the potential barrier at $L_b/a>10$ the deviation in the conductance becomes larger.

Furthermore, we analyze the conductance as a function of the ratio $A_s/A_b$ and of the barrier chemical potential $C_b$, as shown in Fig.~\ref{SOCandvFMismatch}(b). Since the BHZ parameter $A$ modifies the group velocity scale of the bands, varying $A_s/A_b$ provides an effective way to simulate a Fermi-velocity mismatch between the superconducting region and the barrier. At the same time, it can also mimic band-bending effects induced by the coupling to the superconductor, which may locally modify the dispersion of the proximitized barrier modes.

Generically, a Fermi-velocity mismatch reduces the transparency of the barrier--superconductor interface and therefore makes the proximity effect less efficient. In this case, the induced gap $\Delta_{\text{eff}}$ is reduced, weakening electron-hole conversion inside the barrier and the conductance. Conversely, when the velocity mismatch is reduced and the barrier couples more efficiently to the superconductor, $\Delta_{\text{eff}}$ increases, the edge-to-edge coupling is suppressed, and the conductance moves closer to its quantized value. Thus, panel~(b) shows that, in addition to orbital mismatch and finite doping, Fermi-velocity mismatch provides another control knob for tuning the effective coupling between the barrier and the superconductor.

\section{NS conductance in the disordered regime}
In this appendix, we further analyze the disordered QSH/barrier/SC-junction. We first introduce the mean free path, then we examine the disordered induced opening of tunneling channels, and finally we examine the conductance in the long barrier regime, where more than one band is present at the Fermi energy. 

\subsection{Mean free path}
\label{app.effectivediff}

In the main text, we have seen that the effective reflection probability is attenuated exponentially by the presence of a disorder in the barrier, that is
\begin{align}
    R_\text{eff}= \rho_0 \exp(-2W/\xi_\text{eff})\times  \exp(-W/\ell_{\mathrm{mfp}})
\end{align}
with $\ell_{\mathrm{mfp}}$, the mean free path. We estimate $\ell_{\mathrm{mfp}}$ using the Born approximation and taking into account that the uncorrelated disorder potential $U$ is uniformly distributed within $[-\lambda,\lambda]$, with zero mean and variance
$\langle U^2\rangle=\lambda^2/3$. Under these conditions, the corresponding elastic mean free path is given by 
\begin{align}
    \label{Born mean free path}
    \ell_{\mathrm{mfp}}
    =
    \frac{3\hbar v_{F,\mathrm{barrier}}}
    {2\pi\lambda^2 d(E_F)},
\end{align}
where $v_{F,\mathrm{barrier}}$ is the Fermi velocity of the relevant barrier mode and $d(E_F)$ is the density of states per site at the Fermi energy. To fit the numerical curves shown in Fig.~\ref{DisorderConductance}, we use the same values of $\rho_0$ for the clean system ($\lambda=0$) and obtain $\ell_{\mathrm{mfp}}$ extracting the Fermi velocity $v_{F,\mathrm{barrier}}$ and density of states at the Fermi energy $d(E_F)$ from the normal spectrum of the nanoribbon. Although Eq.~\eqref{Born mean free path} is formally valid in the weak-disorder regime, we find a good agreement of the numerical results of the conductance as a function of $W$ over a broad range of disorder strengths, see Fig.~\ref{DisorderConductance}(a). In particular, the model captures both the increase of the conductance with $\lambda$ and the disorder-induced modification of its overall width dependence, while the numerical curves retain sample-specific oscillations. 

\begin{figure}[t!] \centering \includegraphics[width=1\linewidth]{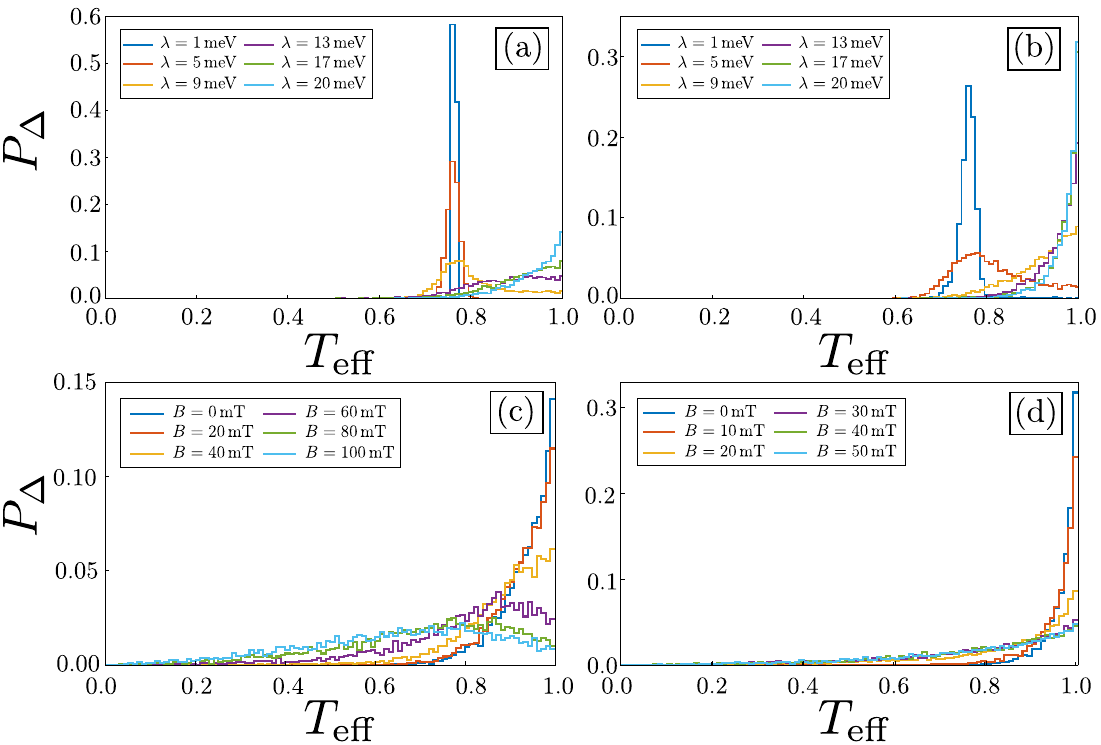} \caption{Binned probability distribution $P_\Delta$ versus the effective transmission $T_{\text{eff}}$, using a bin width $\Delta=0.01$. The height of each bin gives the probability of finding $T_{\text{eff}}$ within that bin. Top row: distributions for different disorder strengths $\lambda$ at (a) $L_b/a$=5 and (b) $L_b/a=20$. Bottom row: distributions for different magnetic fields $B$ at (c) $L_b/a$=5 and (d) $L_b/a=20$. In all plots we average over 5000 disorder configurations.} \label{Statistics} \end{figure}

\subsection{Distribution of the effective transparency}

In this section, we study the distribution function of the effective transmission probability $T_\text{eff}$ and provide an alternative perspective on reflectionless tunneling, commonly referred to as the disorder-induced opening of tunneling channels \cite{Nazarov1994,Beenakker1994}. 

We first consider a clean QSH/barrier/SC junction, for which the zero-bias conductance satisfies $G<4e^2/h$. Using Eq.~(\ref{Andreev conductance}), we find an effective transmission $T_{\mathrm{eff}}<1$ for the single available transport channel. In the presence of disorder, the conductance of the system is the average over many disorder configurations
\begin{align}
\langle G \rangle=\frac{1}{N}\sum_{i=1}^{N}G_i,
\end{align}
where $N$ is the number of disorder realizations and $G_i$ denotes the conductance of realization $i$. Rather than considering only the disorder average, we use Eq.~(\ref{Andreev conductance}) to extract an effective transmission $T_{\mathrm{eff},i}$ from each $G_i$ and construct the corresponding probability distribution $P_{\Delta}(T_{\mathrm{eff}})$. To obtain this distribution, we group the values of $T_{\mathrm{eff},i}$ into intervals of width $\Delta=0.01$, and thus, $P_{\Delta}(T_{\mathrm{eff}})$ denotes the fraction of disorder realizations within each interval. While the clean junction is characterized by a single value $T_{\mathrm{eff}}<1$, disorder redistributes the effective transparencies close to $T_{\mathrm{eff}}=1$ and a rapidly decaying tail toward lower transmissions, as shown in Figs.~\ref{Statistics}(a) and (b). This distribution of the effective transmission then translates to an enhanced average conductance.

Applying a magnetic field introduces a relative phase between the electron and hole paths, suppressing the conductance enhancement that we have discussed in the main text, see Figs.~\ref{Statistics}(c) and (d). 
We can see this effect directly from the distribution of the effective transmission probabilities. As the magnetic field increases, the effective transmission probabilities redistribute towards smaller values, explaining the suppression of the conductance enhancement. In addition, the variance of the distribution increases with larger values of the magnetic field.

\subsection{Long barrier}
\label{App:Long barrier}

In the main text, we have mainly analyzed the backscattering effects of having a barrier $L_b=5a-20a$, hosting one or two bands at the Fermi level. In this section, we include results with a larger $L_b$ and number of bands. 

\begin{figure}[t!] 
\centering \includegraphics[width=1\linewidth]{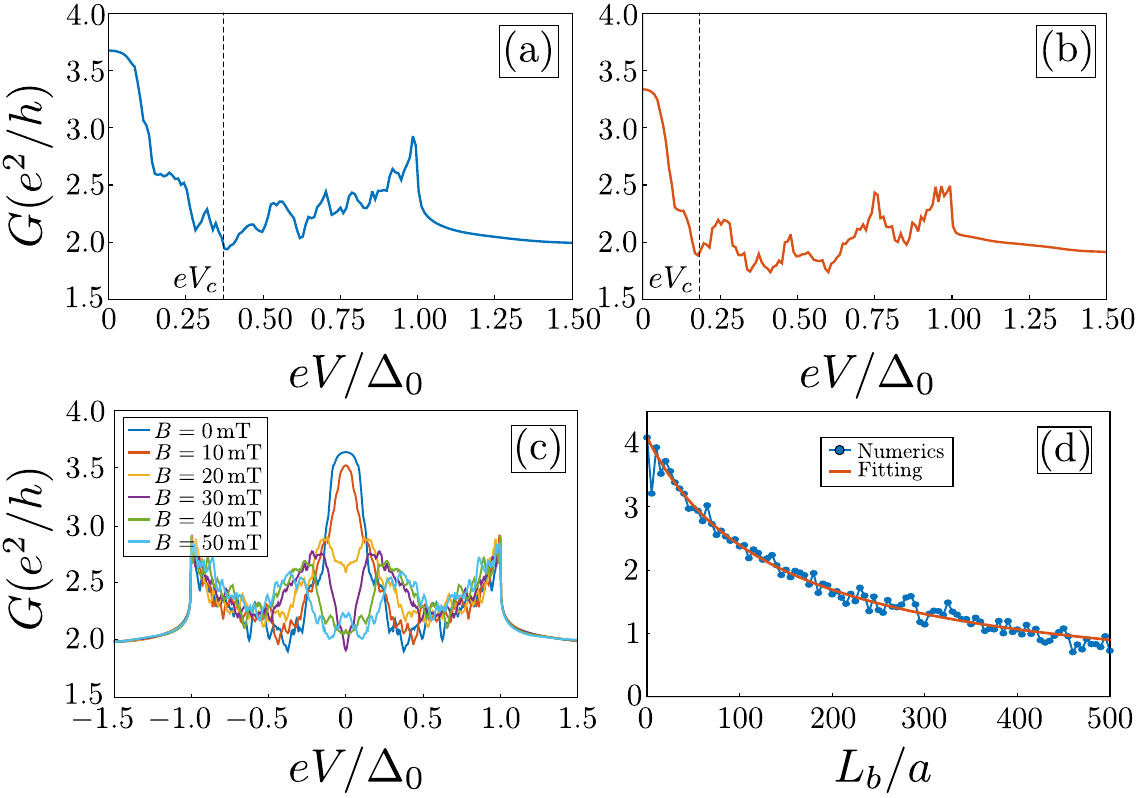} \caption{Conductance $G$ versus bias voltage $eV$ for (a) $L_b/a=20$, (b) $L_b/a=30$ and for (c) $L_b/a=20$ with different magnetic fields $B$. Panel (d): Conductance $G$ versus the length of the barrier $L_b$. The red curve is the phenomenological fitting of Eq.~(\ref{LongLb}) with $\ell/a=280$. In all plots we used $\lambda=20\,$meV and averaged over 100 disorder configurations.} \label{LongLbFigure} 
\end{figure}

First, we study the critical bias voltage at which the phase coherence between the electrons and holes is destroyed. As mentioned in the main text, when the barrier is long enough to permit more than one band at the Fermi energy, the critical bias voltage takes the form $eV_c\simeq \hbar/\tau_{\text{dwell}}$, where $\tau_{\text{dwell}}$ is the dwell time of a quasiparticle in the barrier region, i.e.~the time available to accumulate the relative phase~\cite{Melsen1994}. We estimate the dwell time as the lifetime of a quasiparticle in an open cavity $\tau_{\text{dwell}}=h/NT_{\text{eff}}^c\delta$, where $\delta=1/d(E_F) L_b W$ is the mean level spacing of the barrier region and $N=1$ is the number of escape channels, see Fig.~\ref{Fig1}(a). Here, $T_{\text{eff}}^c$ is the incoherent effective transparency, i.e.~the effective transparency extracted at $V=V_c$ where the interference enhancement is already suppressed. In Figs.~\ref{LongLbFigure}(a,b) we show that the phenomenological estimate for the critical bias voltage agrees well with the numerical calculation.

Furthermore, we examine the behavior of the finite bias conductance for different magnetic fields for a barrier of length $L_b/a=20$, as illustrated in Fig.~\ref{LongLbFigure}(c). For a large disorder strength $\lambda=20\,$meV and zero magnetic field we see a zero-bias conductance peak, similar to Fig.~\ref{DisorderConductance}(c). However, due to the length of the barrier, the conductance near the energy gap $eV=\Delta_0$ does not rise to the full conductance of $G=4e^2/h$, as it does in the case of short barriers. By applying a magnetic field, we see how the peak gradually decreases until it takes the value of the surrounding conductance values.

Finally, we analyze the zero-bias conductance in the regime where the barrier length becomes comparable to the mean free path, $L_b \gtrsim \ell_{\mathrm{mfp}}$. In this regime, we model the average resistance as the sum of two contributions: the ballistic NS-interface resistance, $R_0=h/4e^2$, and the Ohmic resistance associated with propagation through the disordered barrier, $R_{\mathrm{dis}}=(h/2e^2)(L_b/\ell_{\text{Ohm}})$. This leads to the phenomenological expression for the average conductance
\begin{align}
\label{LongLb}
\langle G \rangle
=
\frac{4e^2}{h}
\frac{1}{1+2L_b/\ell_{\text{Ohm}}},
\end{align}
where $\ell_{\text{Ohm}}$ is used as an effective fitting parameter. Eq.~(\ref{LongLb}) provides a good description of the numerical results, as shown in Fig.~\ref{LongLbFigure}(d).

\section{Scattering network model} 
\label{Scattering network calculations} 
In this appendix, we introduce a phenomenological scattering model that captures the essential physics of the interedge coupling and reproduces the results of Sec.~\ref{Andreev bound states and SQI pattern}.

\subsection{Results of the scattering network}
\label{Results of the scattering network}
\begin{figure}[t] \includegraphics[width=0.49\textwidth,height=5cm]{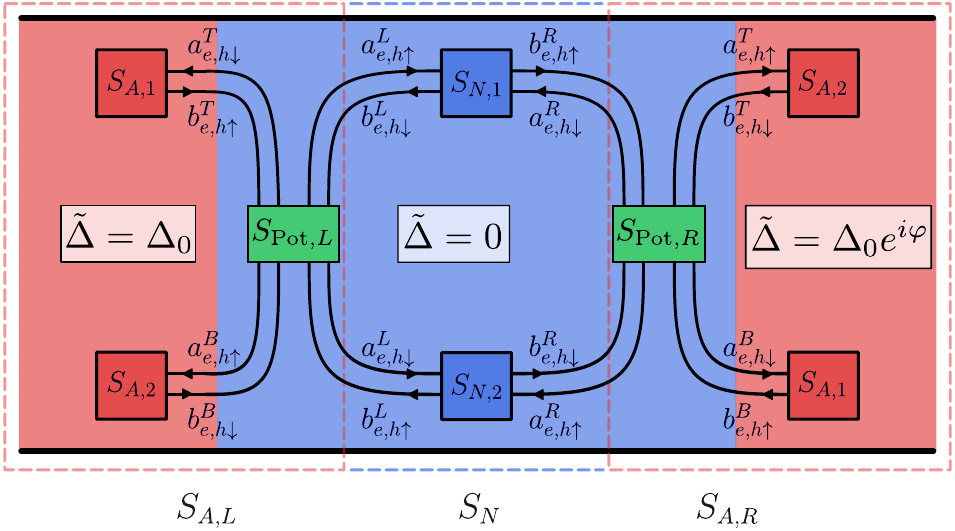} \caption{The phenomenological scattering model of a Josephson junction. The helical edge states in the normal part of the junction can enter the superconductor by local Andreev reflection ($S_A$). Additionally, the top and bottom edges are coupled through $S_{Pot}$, which mimics the scenario displayed in Fig.~\ref{Fig1}(c).} \label{ScatteringJJ} \end{figure}

We can shed some light on the role of the tight-binding model parameters by means of a phenomenological scattering model, which captures the main aspects of the Josephson junction, i.e. the coupling of the helical edges in the normal part and the local Andreev reflection at the NS boundary. The Andreev bound states follow by numerically solving the determinant equation \begin{align} \label{Determinant-Equation} \text{Det}[\mathbb{1}-S_N(E)S_A(E,\varphi)]=0 \end{align} for the energy, where $\mathbb{1}$ is the identity matrix, $S_N$ describes the propagation of the helical edges in the normal part and $S_A$ accounts for the Andreev reflection at the NS interface and the coupling of the edges, which is presented in Fig.~\ref{ScatteringJJ}, see App.~\ref{Appendix A}  for more details. A key observation from the scattering model, is that the effective transmission probability $T_{\text{eff}}$ in Eq.~(\ref{Spot}) controls the size of the gaps. This is expected, as $T_{\text{eff}}$ describes the coupling between the two QSH edges, e.g. by setting $T_{\text{eff}}=1$ we decouple the top and the bottom edge and all gaps close. Furthermore, we find that the phase of a full roundtrip in the normal part $\Psi=2(\alpha_L+\alpha_R)-(\beta_L+\beta_R)+\gamma_t+\gamma_b$, coming from Eqs.~(\ref{SN-Matrix}) and (\ref{Spot}), is directly related to the resonance condition in Eq.~(\ref{Resonance condition}). 
Specifically, the scattering network reveals the following relationships: 
\begin{align} 
\delta_0 &\propto |\cos( \Psi/2)|,\\ \label{ScatteringDeltaPi} \delta_\pi &\propto |\sin( \Psi/2)|.
\end{align} 
In the main text, we showed that $\delta_{0/\pi}$ depend on the microscopic parameters as
\begin{align}
    \delta_0&\propto\left|\cos\left(C_n/E_{T,n}+C_b/E_{T,W}\right)\right|,\\
    \delta_\pi&\propto\left|\sin\left(C_n/E_{T,n}+C_b/E_{T,W}\right)\right|.
\end{align}
By relating the phase of a full roundtrip $\Psi$ to the microscopic parameters we obtain $\Psi=2(C_n/E_{T,n}+C_b/E_{T,W})$, similar to Eq.~(\ref{Resonance condition}) in the main text.

\begin{figure*}[t!] \centering 
\includegraphics[width=1\linewidth]{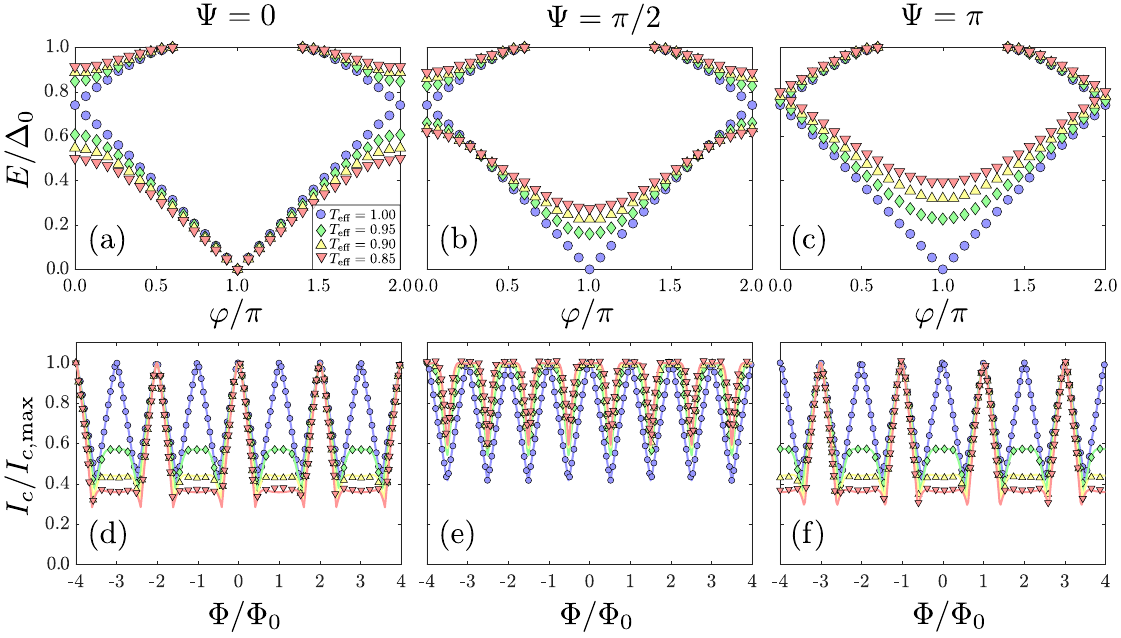} \caption{The Andreev bound states (a)-(c) and the corresponding SQI pattern (d)-(f) obtained from the scattering network for different values of the dynamical phase $\Psi$. Each panel presents the ABS or SQI pattern for different values of the coupling strength $T_{\text{eff},L}=T_{\text{eff},R}=T_{\text{eff}}$.} \label{GapsAndLobes} \end{figure*}

With the help of the scattering network we can also explore the effects of edge coupling in the short and long junction limit, without resorting to computationally intensive tight-binding calculations. In both cases, we can derive an exact result for the energy-phase relation. For the short junction limit $L\ll \xi_s$ we obtain
\begin{align} 
\label{EnergyPhaseShort} E&=\pm \Delta_0 \sqrt{1-T\sin^2(\varphi/2)}, \\ \label{transmission} T&=\frac{ T_{\text{eff},L}T_{\text{eff},R}}{1+R_{\text{eff},L}R_{\text{eff},R}-2\sqrt{R_{\text{eff},L}R_{\text{eff},R}}\cos\Psi)}, 
\end{align} 
which is reminiscent of the energy-phase relation of a trivial Josephson junction with an altered transmission~\cite{Beenakker1991a}. A striking feature of the short junction regime is that the interedge coupling only opens a gap at $\varphi=\pi$, while the gap at $\varphi=0$ remains closed. This is in stark contrast to the intermediate and the long junction regime, as we will see shortly. Furthermore, when the left and right barrier exhibit the same effective transmission probabilities $T_{\text{eff},L}=T_{\text{eff},R}$, we find that the system behaves like a  Fabry-P\'{e}rot interferometer. When $\Psi=2\pi n$, with $n\in\mathbb{Z}$, the transparency $T$ becomes exactly $1$, regardless of the coupling strength between the edges, as we can see from Eq.~(\ref{transmission}).

Contrary to the short junction limit, the long junction captures the Fabry-P\'{e}rot oscillations of $\textit{both}$ gaps, making this scenario similar to the intermediate junction. In this regime, the dominant energy scale is the superconducting coherence length and the energy-phase relation reads as 
\begin{align}
\label{eq:long-junction-limit}
E_{m,\pm}
    &=E_T\left[m\pi\pm\tfrac{1}{2}\arccos\chi\right],
    \\
\chi
    &=\frac{
        T_{\mathrm{eff}}^2\cos\varphi
        +4(T_{\mathrm{eff}}-1)\cos\Psi
    }{(2-T_{\mathrm{eff}})^2},
\end{align}
where $E_T=\Delta_0 \xi_s/L$ is the Thouless energy and $m\in\mathbb{Z}$ is the branch index. Furthermore, we assumed $T_{\text{eff,L}}=T_{\text{eff,R}}=T_{\text{eff}}$ for simplicity. Here, we cannot express Eq.~(\ref{eq:long-junction-limit}) in terms of a transmission function $T$ as in Eq.~(\ref{EnergyPhaseShort}), but we can relate it to the gap amplitudes $\delta_0$ and $\delta_\pi$, which are also measures of the transmission. Specifically, we find that $\delta_{\pi (0)}=0$ for $\Psi=2\pi n\ ([2n+1]\pi)$, for $n\in\mathbb{Z}$, regardless of the edge coupling strength, which is precisely the observation shown in Fig.~\ref{GapsAndLobes}(a)-(c). In other words, for certain values of $\varphi$ and $\Psi$, the energy $E$ becomes independent of $T_{\text{eff}}$, and the system behaves as if the edges were decoupled since all values for $T_{\text{eff}}$ are equal~\footnote{Note, that $T_{\text{eff}}=0$ is a special case since it eliminates the $\varphi$-dependence. Physically it means that if we cut the connection between the normal and the superconducting part, we also cannot have Andreev bound states.}. The scattering network also captures the connection between the gaps in the ABS and the reduction of lobes in the SQI, as pointed out in Sec.~\ref{Andreev bound states and SQI pattern}. The Josephson current at finite temperature $T=55\,$mK can be calculated from $S_N$ and $S_A$ as \begin{align} \label{Supercurrent} I(\varphi)=-kT \frac{2e}{\hbar}\frac{d}{d\varphi}\sum_{n=0}^{\infty}\text{ln}\ \text{Det}[\mathbb{1}-S_{N}(i\omega_n)S_A(i\omega_n,\varphi)], \end{align} where $\omega_n=\pi k T(2n+1)$ are the Matsubara frequencies and $k$ is the Boltzmann constant. In Fig.~\ref{GapsAndLobes} (d)-(f) we show the SQI pattern and observe that the reduction of the lobes is tied to the opening of the gaps in the ABS, shown in Fig.~\ref{GapsAndLobes} (a)-(c). More specifically, we observe that the case of $\delta_{\pi (0)} $=0 and $\delta_{0 (\pi)}\neq 0$ yields a situation where the odd (even) lobes are maximally suppressed compared to the even (odd) lobes, which is precisely the same observation as in Subsec.~\ref{Andreev bound states and SQI pattern}. Furthermore, we find a scenario where both gaps in the ABS are open and the reduction of the even and odd lobes is equal, effectively restoring the SQUID pattern even in the presence of coupled edges, as shown in Fig.~\ref{GapsAndLobes} (f). 

We can make further use of the scattering network and investigate the effects of disorder on transport properties, specifically on the SQI pattern. In our system, time-reversal symmetry protects the helical edges from disorder that respects this symmetry. However, such disorder may still affect the coupling strength between the edges and the dynamical phase that particles acquire during interedge backscattering. To this end, we randomly assign values to the effective transmission probability $T_{\text{eff},L/R}\sim \mathcal{U}[0,1]$ and to the phase $\Psi\sim \mathcal{U} [0, 2\pi)$ for the left ($L$) and right ($R$) interfaces, respectively. Here, $x\sim \mathcal{U}[a,b)$ means that $x$ is randomly drawn from a uniform distribution in the interval $[a,b)$.  Then, we extract the critical current $\text{Max}_{\varphi}|I(\varphi)|$ and average it over $1000$ disorder configurations  \begin{align} I_c=\langle \text{Max}_{\varphi}|I(\varphi)|\rangle_{\text{disorder}}. \end{align} Performing the calculation reveals, that the even–odd lobe modulation almost vanishes in the presence of disorder, as shown in Fig~\ref{SQIDisorder}.

\subsection{Technical details of the scattering network model}
\label{Appendix A}
In this subsection, we will provide a detailed explanation of how to obtain the scattering matrices $S_N$ and $S_A$ from Fig.~\ref{ScatteringJJ}. First, we note that we can write $S_A$ and $S_N$ in the following basis: 
\begin{equation}
\label{SAandSN}
    \begin{pmatrix}
        a^{L} \\
        a^{R}
        \end{pmatrix}=
        S_{A}
    \begin{pmatrix}
        b^{L}\\
        b^{R}
    \end{pmatrix}, \ \ \ \ \begin{pmatrix}
        b^{L} \\
        b^{R}
    \end{pmatrix}=S_{N} \begin{pmatrix}
        a^{L}\\
        a^{R}
    \end{pmatrix}
\end{equation}
with:
\begin{align}
\label{Basis1}
a^{L/R}&=\left(a_{e,\uparrow}^{L/R},a_{e,\downarrow}^{L/R},a_{h,\uparrow}^{L/R},a_{h,\downarrow}^{L/R} \right)^{T}\\
\label{Basis2}
 b^{L/R}&=\left(b_{e,\uparrow}^{L/R},b_{e,\downarrow}^{L/R},b_{h,\uparrow}^{L/R},b_{h,\downarrow}^{L/R} \right)^{T}.
\end{align}
$e$ ($h$) indicates electrons (holes) and $\uparrow$ ($\downarrow$) represents spin up (spin down) particles. $L$ and $R$ denote the left and right-hand-side of the junction as pictured in Fig.~\ref{ScatteringJJ}.\\
$S_{N}$ gives the relation between the incoming ($a^{L}$, $a^{R}$) and outgoing ($b^{L}$, $b^{R}$) modes relative to the central part of the junction. Note, that $S_{N,1}$ and $S_{N,2}$ in Fig.~\ref{ScatteringJJ} make up $S_{N}$ upon expressing the two scattering matrices in the basis of Eqs.~(\ref{Basis1}) and (\ref{Basis2}). In the central region of the junction, spin-flipping processes are suppressed due to the topological protection of the edge states even in the presence of potential barriers and disorder. Also, the charge of the mode cannot change far away from the NS boundary. Thus, $S_N$ consists solely of matrix elements connecting modes of the same spin and same charge. Nonetheless, the propagating edges can accumulate a Thouless phase $\chi_{t/b}=E/E_{T,t/b}$, a static phase $\gamma_{t/b}$ and a magnetic flux $\tilde{\Phi}=\pi\Phi/\Phi_0=(e/\hbar)\int_0^{L} A_y|_{x=\text{W}} dy$ with ${\bf A}=-By\bm{e}_x$. $E_{T,t/b}$ is the top/bottom Thouless energy. Thus, $S_{N}$ takes the following form:
\begin{equation}
\label{SN-Matrix}
    S_N=\left( \begin{array}{cccc}
         \mathbb{0}_{4\times 4} & S_{N,LR}  \\
         S_{N,RL}& \mathbb{0}_{4\times 4} 
    \end{array}
    \right),
\end{equation}
with $\mathbb{0}_{4\times 4}$ being the $4\times 4$ zero matrix and
\begin{equation}
S_{N,LR}=\left(
\begin{array}{cccc}
  e^{i (\zeta_b+\tilde{\Phi}/2)} & 0 & 0 & 0 \\
  0 & e^{i (\zeta_t-\tilde{\Phi}/2)} & 0 & 0 \\
  0 & 0 & e^{i( \zeta_b-\tilde{\Phi}/2)} & 0 \\
  0 & 0 & 0 & e^{i (\zeta_t+\tilde{\Phi}/2)} \\
 \end{array}
\right) \nonumber
\end{equation}

\begin{equation}
S_{N,RL}=\left(
\begin{array}{cccc}
 e^{i (\zeta_t+\tilde{\Phi}/2)} & 0 & 0 & 0  \\
 0 & e^{i (\zeta_b-\tilde{\Phi}/2)} & 0 & 0  \\
 0 & 0 & e^{i (\zeta_t-\tilde{\Phi}/2 )} & 0 \\
 0 & 0 & 0 & e^{i (\zeta_b+\tilde{\Phi}/2)}\\
\end{array}
\right). \nonumber
\end{equation}
where $\zeta_{t/b}=\chi_{t/b}+\gamma_{t/b}$. To describe the scattering processes at the NS interface, we express the $S$-matrix as follows
\begin{align}
\label{SA}
    S_{A}=\begin{pmatrix}
        S_{A,L} & \mathbb{0}_{4\times 4} \\
        \mathbb{0}_{4 \times 4} & S_{A,R}
    \end{pmatrix},
\end{align}

\begin{figure}
\includegraphics[width=3.in]{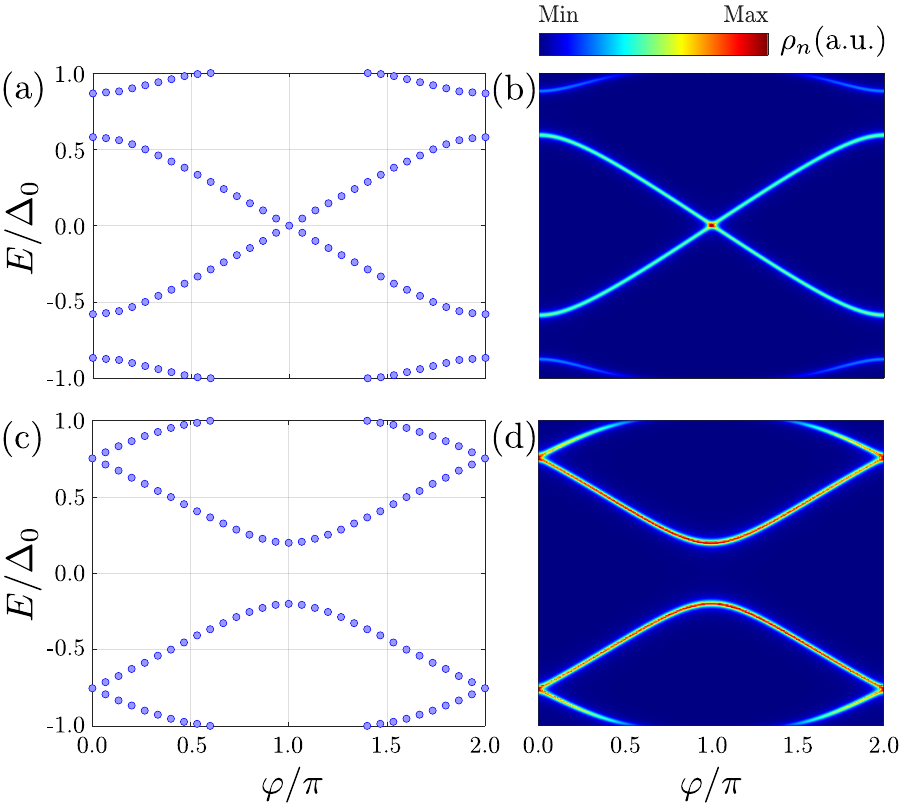}
  \caption{Energy $E$ versus superconducting phase difference $\varphi$ from the scattering model with (a) $\Psi=0$ and (c) $\Psi=\pi$. Spectral density $\rho_n$ versus $E$ and $\varphi$ from the tight-binding model for (b) $\delta_0\neq0$, $\delta_\pi=0$ and (d) $\delta_0=0$, $\delta_\pi\neq0$.}
\label{ComparisonScatteringTB}
\end{figure}

where $S_{A,L}$ and $S_{A,R}$ describe the left and right boundary, respectively. $S_{A,L/R}$ encompasses three scattering matrices, namely $S_{A,1}$, $S_{A,2}$ and $S_{Pot}$, as Fig.~\ref{ScatteringJJ} shows. To obtain $S_{A,L/R}$ from the other matrices, we have to combine them in such a way that we eliminate the modes which are not used to define $S_{A}$ in Eq.~(\ref{SAandSN}), that is we eliminate all the modes which are not given in Eqs.~(\ref{Basis1}) and (\ref{Basis2}). To be more precise, we first write down the six systems of equations that emerge at the left boundary (the right boundary follows suit) and obtain from Fig.~\ref{ScatteringJJ}:
\begin{align}
\label{SOE 1}
    \begin{pmatrix}
        b_{e, \uparrow}^{T} \\
        b_{h, \uparrow}^{T}
            \end{pmatrix}
            &=S_{A,1} \begin{pmatrix}
            a_{e,\downarrow}^{T} \\
            a_{h,\downarrow}^{T}
        \end{pmatrix}, \ \ \ \begin{pmatrix}
        b_{e, \downarrow}^{B} \\
        b_{h, \downarrow}^{B}
            \end{pmatrix}
            =S_{A,2} \begin{pmatrix}
            a_{e,\uparrow}^{B} \\
            a_{h,\uparrow}^{B}
        \end{pmatrix}, \\
        \begin{pmatrix}
        a_{e, \uparrow}^{B} \\
        a_{e, \uparrow}^{L}
            \end{pmatrix}
            &=S_{Pot} \begin{pmatrix}
            b_{e,\uparrow}^{L} \\
            b_{e,\uparrow}^{T}
        \end{pmatrix}, \ \ \begin{pmatrix}
        a_{e, \downarrow}^{T} \\
        a_{e, \downarrow}^{L}
            \end{pmatrix}
            =S_{Pot} \begin{pmatrix}
            b_{e,\downarrow}^{L} \\
            b_{e,\downarrow}^{B}
        \end{pmatrix},\\
        \label{SOE 2}
        \begin{pmatrix}
        a_{h, \uparrow}^{B} \\
        a_{h, \uparrow}^{L}
            \end{pmatrix}
            &=S_{Pot}^{*} \begin{pmatrix}
            b_{h,\uparrow}^{L} \\
            b_{h,\uparrow}^{T}
        \end{pmatrix}, \ \ \begin{pmatrix}
        a_{h, \downarrow}^{T} \\
        a_{h, \downarrow}^{L}
            \end{pmatrix}
            =S_{Pot}^{*} \begin{pmatrix}
            b_{h,\downarrow}^{L} \\
            b_{h,\downarrow}^{B}
        \end{pmatrix},
\end{align}
where 
\begin{align}
\label{SA-Matrix}
S_{A,1}&=\begin{pmatrix}
    0 & r_{eh}^{\uparrow \downarrow}\\
    r_{he}^{\downarrow \uparrow} & 0
\end{pmatrix}, \ \ S_{A,2}=\begin{pmatrix}
    0 & r_{eh}^{\downarrow \uparrow}\\
    r_{he}^{\uparrow \downarrow} & 0
\end{pmatrix}, \ \ \\
\label{Spot}
S_{\text{Pot},L/R}&=
\begin{pmatrix}
e^{ i\alpha_{L/R}} \sqrt{T_{\text{eff}}} & e^{i \beta_{L/R}}\sqrt{1-T_{\text{eff}}}\\
- e^{i(2\alpha_{L/R}-\beta_{L/R})}\sqrt{1-T_{\text{eff}}} & e^{i\alpha_{L/R}}\sqrt{T_{\text{eff}}}
\end{pmatrix}.
\end{align} 
$S_{A,1/2}$ describes the reflection of an incoming electron into an outgoing hole with flipped spin and vice versa, with the reflection amplitudes~\cite{Nazarov2009}:
\begin{align}
\nonumber
    r^{he}_{\downarrow \uparrow}(E)&=(r^{eh}_{\downarrow \uparrow}(-E))^*= \left(\frac{E}{\Tilde{\Delta}}-i\frac{\sqrt{|\Tilde{\Delta}|^2-E^2}}{\Tilde{\Delta}} \right)\\
    \nonumber
    r^{he}_{\uparrow \downarrow}(E)&=(r^{eh}_{\uparrow \downarrow}(-E))^*=- \left(\frac{E}{\Tilde{\Delta}}-i\frac{\sqrt{|\Tilde{\Delta}|^2-E^2}}{\Tilde{\Delta}} \right),
\end{align}
where $\Tilde{\Delta}=\Delta_0$ for the left and $\Tilde{\Delta}=\Delta_0 e^{i\varphi}$ for the right superconductor, as depicted in Fig.~\ref{ScatteringJJ}.  
$S_{\text{Pot},L/R}$ accounts for the coupling of the edges on the left/right side. On one hand, $T_{\text{eff}}$ represents the probability that a mode stays on a given edge. On the other hand, $1-T_{\text{eff}}$ accounts for a change of edges. $\alpha_{L/R}$ and $\beta_{L/R}$ are phases acquired by the mode after scattering with the potential barrier. Furthermore, $S_{\text{Pot},L/R}$ respects time-reversal symmetry and thus it cannot mix modes of different spin species. 

We can now combine the Eqs.~(\ref{SOE 1})-(\ref{SOE 2}) by eliminating the modes $b_{e,h \uparrow}^{T}$, $b_{e,h \downarrow}^{B}$, $a_{e,h \uparrow}^{B}$ and $a_{e,h \uparrow}^{T}$. Thus, we are left with $a_{e,h \uparrow}^{L}$, $a_{e, h \downarrow}^{L}$, $b_{e, h \uparrow}^{L}$ and $b_{e, h \downarrow}^{L}$, which form the $L$ part of the basis given in Eqs.~(\ref{Basis1}) and (\ref{Basis2}). The matrix we obtain from the combination is $S_{A,L}$ if written as $a^L=S_{A,L}b^L$. We will not give the explicit form of $S_{A,L}$, for we combine the systems of equations numerically. In the same way, we obtain $S_{A,R}$ from the right side of Fig.~\ref{ScatteringJJ} and thus $S_A$ with Eq.~(\ref{SA}).

By tuning the parameters of the scattering model, we can mimic the tight-binding calculations to a good degree, as demonstrated in Fig.~\ref{ComparisonScatteringTB}.

\end{document}